\documentclass[final]{IEEEtran}
\newlength \figwidth
\usepackage{cite}
\ifCLASSINFOpdf
  \usepackage[pdftex]{graphicx}
	\usepackage{epstopdf}
  \graphicspath{{../Figures/}}
  \DeclareGraphicsExtensions{.eps}
\else
  \usepackage[dvips,draft]{graphicx}
\fi
\usepackage[cmex10]{amsmath}
\usepackage{amssymb}
\usepackage{pifont}
\usepackage{amsthm}
\usepackage{url}
\usepackage{dsfont}
\usepackage{tabulary}
\usepackage{multirow}
\usepackage{color}

\usepackage{color}
\usepackage{times}
\usepackage{verbatim}
\usepackage{floatflt}
\usepackage{enumerate}
\usepackage{array}
\usepackage{multicol,afterpage,wrapfig} 
\usepackage{tikz}
\usepackage[noend]{algpseudocode}
\makeatletter
\def\BState{\State\hskip-\ALG@thistlm}
\makeatother
\usepackage{amsmath,amssymb,eucal}
\usepackage[utf8]{inputenc}
\usepackage{epsfig}
\usepackage{exscale}
\usepackage{latexsym}
\usepackage{verbatim}
\usepackage{amsfonts}
\usepackage{multirow}
\usepackage{cite}
\usepackage{balance}
\usepackage{color}
\usepackage{transparent}
\usepackage{import}
\usepackage{mathtools}
\usepackage{tikz}
\usetikzlibrary{automata,arrows,positioning,calc}
\usepackage{bbm}
\usepackage[printonlyused,withpage]{acronym}
\usepackage{tabu}
\usepackage{longtable}
\usepackage{balance}
\usepackage{footnote}
\usepackage{tablefootnote}
\usepackage{adjustbox}
\usepackage{multirow}
\usepackage{pifont}
\newcommand{\cmark}{\ding{51}}%
\newcommand{\xmark}{\ding{55}}%

\def\BibTeX{{\rm B\kern-.05em{\sc i\kern-.025em b}\kern-.08em
    T\kern-.1667em\lower.7ex\hbox{E}\kern-.125emX}}
    
\renewcommand{\IEEEQED}{\IEEEQEDopen}

\newcommand*\xbar[1]{%
  \hbox{%
    \vbox{%
      \hrule height 0.5pt 
      \kern0.36ex
      \hbox{%
        \kern-0.12em
        \ensuremath{#1}%
        \kern-0.12em
      }%
    }%
  }%
}

\newfont{\bbb}{msbm10 scaled 500}

\newfont{\bb}{msbm10 scaled 1100}

\newcommand{\executeiffilenewer}[3]{%
\ifnum\pdfstrcmp{\pdffilemoddate{#1}}%
{\pdffilemoddate{#2}}>0%
{\immediate\write18{#3}}\fi%
}

\usetikzlibrary{arrows,calc}
\allowdisplaybreaks

\IEEEoverridecommandlockouts
\begin{document}
\pagenumbering{gobble}

\newtheorem{Theorem}{\bf Theorem}
\newtheorem{Corollary}{\bf Corollary}
\newtheorem{Remark}{\bf Remark}
\newtheorem{Lemma}{\bf Lemma}
\newtheorem{Proposition}{\bf Proposition}
\newtheorem{Assumption}{\bf Assumption}
\newtheorem{Definition}{\bf Definition}
\title{Understanding UAV Cellular Communications: From Existing Networks to Massive MIMO}
\author{\IEEEauthorblockN{{Giovanni~Geraci, Adrian~Garcia-Rodriguez, Lorenzo~Galati~Giordano, David~L\'{o}pez-P\'{e}rez, and Emil Bj\"{o}rnson}}
\thanks{G.~Geraci, A.~Garcia-Rodriguez, L.~Galati~Giordano, and D.~L\'{o}pez-P\'{e}rez are with Nokia Bell Labs, Ireland. E.~Bj\"{o}rnson is with Link\"{o}ping University, Sweden. Corresponding author's email: dr.giovanni.geraci@gmail.com. The material in this paper will in part be presented at IEEE ICC 2018 \cite{GerGarGalICC2018}.}}
\maketitle
\thispagestyle{empty}
\begin{abstract}
The purpose of this article is to bestow the reader with a timely study of UAV cellular communications, bridging the gap between the 3GPP standardization status quo and the more forward-looking research. Special emphasis is placed on the downlink command and control (C\&C) channel to aerial users, whose reliability is deemed of paramount technological importance for the commercial success of UAV cellular communications. Through a realistic side-by-side comparison of two network deployments -- a present-day cellular infrastructure versus a next-generation massive MIMO system -- a plurality of key facts are cast light upon, with the three main ones summarized as follows: 
\emph{(i)} UAV cell selection is essentially driven by the secondary lobes of a base station's radiation pattern, causing UAVs to associate to far-flung cells; 
\emph{(ii)} over a 10~MHz bandwidth, and for UAV heights of up to 300~m, massive MIMO networks can support 100~kbps C\&C channels  in 74\% of the cases when the uplink pilots for channel estimation are reused among base station sites, and in 96\% of the cases without pilot reuse across the network;
\emph{(iii)} supporting UAV C\&C channels can considerably affect the performance of ground users on account of severe pilot contamination, unless suitable power control policies are in place.
\end{abstract}

\IEEEpeerreviewmaketitle
\section{Introduction}


The latest proliferation of unmanned aerial vehicles (UAVs) -- more commonly referred to as \emph{drones} -- is generating thrill, charm, and engagement in the public and private domains alike. Highly mobile UAVs are regarded as the best candidates to automate and ease emergency search-and-rescue missions, crowd management and surveillance, as well as weather and traffic monitoring \cite{NewAmerica2015,ValVac2005}. Their recently reduced cost also makes drones suitable for less critical applications such as parcel delivery and video streaming of breathtaking landscapes. All but unheard of until just recently, drones are now envisioned to shape the future of technology, providing a useful, trustworthy, and safe solution for human endeavors \cite{USDeptTransp,MozSaaBen2018}. Moreover, a rapid and vast growth in the UAV business will likely open attractive vertical markets in the telecommunications industry, bringing new revenue opportunities for mobile network vendors and operators. 
 
\subsection{Background and Motivation}

For the above technological and commercial visions to turn into reality, UAV users will require control and connectivity over a wireless network. Solutions are urgently needed to transfer large amounts of data, e.g., video streaming generated by high-resolution cameras, from aerial users to ground base stations (BSs) \cite{ChaLar2017, ZenLyuZha2018, 3GPP36777}. More importantly, reliable command and control (C\&C) channels to the UAVs are required to safely operate these vehicles remotely and beyond the visual line-of-sight (LoS) constraints currently enforced by regulatory bodies \cite{ZenZhaLim2016}. In this setup, terrestrial cellular networks are well positioned to serve UAV users flying up to an altitude of few hundred meters.
 
Although connecting UAVs through cellular technologies has key potential advantages -- such as enabling connectivity via existing network infrastructure and spectrum resources -- it also involves important challenges. Indeed, UAVs may undergo radio propagation characteristics that are profoundly different from those encountered by conventional ground users (GUEs). As they can move in three dimensions, UAVs could be placed in locations considerably above ground, experiencing a larger distance to the radio horizon and favorable LoS propagation conditions with a vast number of BSs \cite{KhaGuvMat2018}. As a result, a UAV transmitting uplink (UL) information to its serving BS could create significant interference to a plurality of neighboring BSs, receiving both GUE and UAV UL transmissions. Conversely, BSs communicating with their GUEs or UAVs in downlink (DL) could severely disrupt the DL of a UAV associated to a neighboring BS \cite{HayYanMuz2016,BerChiPol2016}.

In light of these challenges, and with the aim of integrating UAV communications in current cellular networks, the Third Generation Partnership Project (3GPP) has been gathering key industrial players to collaborate on a \emph{study item} on enhanced LTE support for aerial vehicles. Notably, such joint effort -- that has just reached its conclusion, and that will soon be followed by the corresponding \emph{work item} -- has produced systematic measurements and accurate modeling of UAV-to-ground channels \cite{3GPP36777}. Furthermore, it has defined the various UAV link types along with their respective features and minimum requirements, as summarized in Table~\ref{table:Link_types}. The remarkable industrial involvement in UAV cellular communications standardization \cite{Qualcomm2017,RP170779,R11713831,R11714071,R11717287,R11718267,R11717873}, together with the concurrent theoretical investigations being undertaken in academia \cite{WanJiaHan2017,ChaDanLar2017,AzaRosPol2017,GalKibSilva2017,7536859,KalBor2017}, prompt us to bridge the gap and provide a compelling study that follows the most recent 3GPP recommendations and sets the trend for present-and-forward-looking research.

\subsection{Approach and Summary of Results}

In this article, we aim at advancing the understanding of UAV cellular communications, paying particular attention to the performance of the UAV DL C\&C channel, for which a minimum requirement of 100~kbps has been defined \cite{3GPP36777}. Such focus is motivated by the pursuit of mobile operators for new market shares. Unlike UAV payload traffic, potentially similar to -- and in lower amount than -- current data traffic, handling C\&C traffic could yield additional profits that may justify investments to upgrade the network infrastructure. In light of this prospect, we extend our study to two network architectures, both operating on a 10~MHz bandwidth: \emph{(i)} a traditional network with sectorized BSs operating in \emph{single-user} mode (i.e., one user scheduled per frequency-time resource at a time) -- representing most existing cellular deployments; and \emph{(ii)} a massive MIMO cellular network operating in \emph{multi-user} mode (i.e., multiple users scheduled per frequency-time resource) with digital beamforming capabilities -- exemplifying next-generation deployments.

For these practical scenarios, we closely examine how the height of a UAV user affects its cell selection process and its performance. We also evaluate the increased reliability that can be achieved for the UAV C\&C channel through massive MIMO, and we quantify what the presence of UAVs entails for the performance of conventional GUEs. While we refer the reader to the body of the article for a comprehensive collection of results and discussions, the following list serves as a preview of the most important takeaways of our study:
\begin{itemize}
\item 
For UAV heights of 75~m and above, due to an almost free-space propagation, cell selection in existing networks is mostly driven by the secondary lobes of each BS's antenna pattern, rather than by the path loss difference among BSs. Hence, UAVs do not generally associate to BSs located nearby but to those found farther away.
\item 
In current single-user mode networks designed for GUEs, UAVs taking off or landing at 1.5~m achieve the C\&C target rate of 100~kbps in 87\% of the cases. However, because of strong LoS interference received from a plurality of visible cells, such reliability decreases to 35\% at 50~m, and to a mere 2\% and 1\% at 150~m and 300~m, respectively.
\item 
Multi-user massive MIMO can support a 100~kbps C\&C channel for UAV heights up to 300~m with 74\% and 96\% reliability, respectively with and without pilot reuse and contamination. This is due to a mitigated interference, a stronger carrier signal, and a spatial multiplexing gain.
\item 
The presence of UAVs can significantly degrade the performance attained by GUEs with massive MIMO. UL power control policies are required to protect GUEs whose pilot signals are otherwise vulnerable to severe contamination from UAV-generated overlapping pilots.
\end{itemize}

\begin{table}[t]
\centering
\caption{Taxonomy of UAV link types}
\label{table:Link_types}
\renewcommand{\arraystretch}{1.5}
\begin{tabular}{|c|c|c|c|}\hline
\textbf{} & \multicolumn{1}{c|}{\textbf{Link Type}} & \multicolumn{1}{c|}{\textbf{Data Rate}} & \multicolumn{1}{c|}{\textbf{Critical?}} \\\hline
\multirow{3}{*}{DL} & {Synchronization (PSS/SSS)} & \multirow{2}{*}{N/A} & \cmark \\\cline{2-2}\cline{4-4}
& {Radio control (PDCCH)} &  & \cmark \\\cline{2-4}
 & \multicolumn{1}{c|}{{C\&C}} & \multicolumn{1}{c|}{60-100~kbps} & \cmark \\\cline{1-4}
\multirow{2}{*}{UL} & \multicolumn{1}{c|}{C\&C} & \multicolumn{1}{c|}{60-100~kbps} & \cmark \\\cline{2-4}
 & \multicolumn{1}{c|}{Payload} & \multicolumn{1}{c|}{Best effort} & \xmark \\\hline
\end{tabular}
\end{table}
\section{3GPP System Setup}
\label{sec:System_Model}

In this section, we introduce the 3GPP network topology and channel model employed in this paper. Further details on the specific parameters used for our studies are given in Table~\ref{table:parameters}.

\subsection{Cellular Network Topology}

\begin{figure*}[!t]
\centering
\includegraphics[width=1.9\columnwidth]{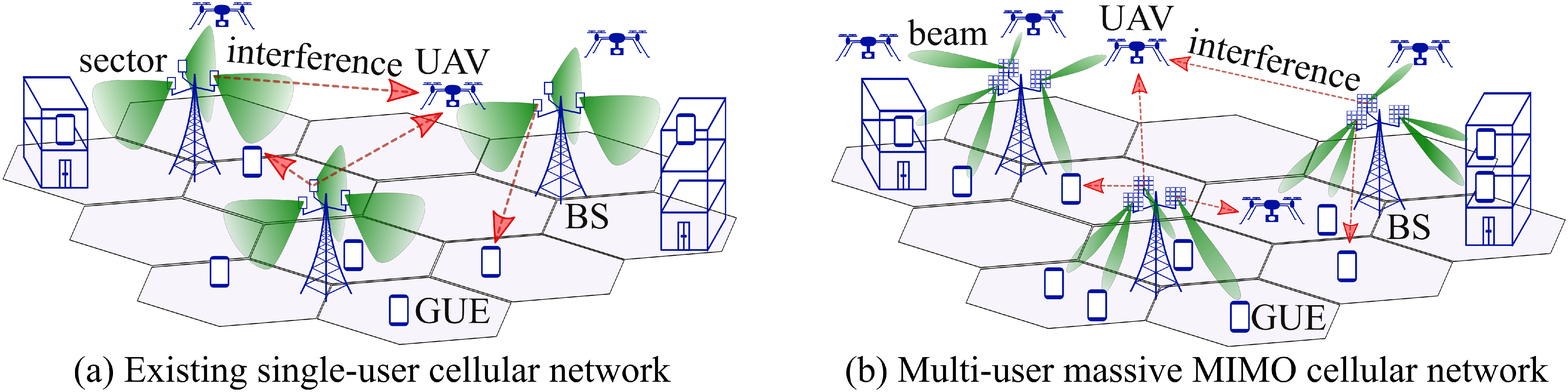}
\caption{Illustration of two examples of cellular infrastructure to support both ground and UAV cellular communications. In (a) -- similarly to most existing networks -- BSs are equipped with a vertical antenna panel to cover a cellular sector and serve a single user on each PRB, potentially generating strong interference towards neighboring users. In (b) -- exemplifying next-generation networks -- BSs are equipped with massive MIMO arrays and serve multiple users on each PRB through beamforming, also increasing the useful signal power at each served user and mitigating the interference towards neighboring users.}
\label{fig:Network}
\end{figure*}

We consider the DL of a traditional cellular network (designed for GUEs) as illustrated in Fig.~\ref{fig:Network}, where BSs are deployed on a hexagonal layout to communicate with their respective sets of connected users. A deployment site is comprised of three co-located BSs, each covering one sector spanning an angular interval of $120^{\circ}$. Unlike conventional cellular networks, the network under consideration serves both GUEs and UAVs, e.g., providing the former with DL data streams and the latter with C\&C information. In what follows, the term \emph{users} denotes both GUEs and UAVs. GUEs are located both outdoor (at a height of 1.5~m) and indoor in buildings that consist of several floors. UAVs are located outdoor at variable heights between 1.5~m, which characterizes their height during take off and landing, and 300~m, which represents their maximum cruising altitude with cellular service. All deployment features comply with the ones specified by the 3GPP in \cite{3GPP36777}.

We denote by $\mathcal{B}$  the set of cellular BSs, and assume that all BSs transmit with total power $P_{\textrm{B}}$. The transmission power per time-frequency physical resource block (PRB) is given by $P_{\textrm{b}}=P_{\textrm{B}}/F$, where $F$ denotes the total number of PRBs. Users associate to the BS that provides the largest reference signal received power (RSRP) across the whole communication band. All BSs are equipped with $N_{\mathrm{a}}$ antennas, and we assume all users to be equipped with a single antenna. We denote as $\mathcal{K}_b$ the set of users served by BS $b$ on a given PRB, and by $K_b$ its cardinality. While the total number of associated users is determined by their density and distribution, the set $\mathcal{K}_b$ can be chosen adaptively by BS $b$ through scheduling operations. In this regard, we identify two cases: the one where $K_b=1$ (\emph{single-user} mode operations) and the one where $K_b \geq 1$ (\emph{multi-user} operations through spatial multiplexing). These two cases are illustrated in Fig.~\ref{fig:Network}(a) and Fig.~\ref{fig:Network}(b), and will be addressed in Sec.~\ref{sec:SU-MIMO} and Sec.~\ref{sec:MU-MIMO}, respectively.

\subsection{3D Propagation Channel Model}

We adopt the newly released 3GPP channel model for evaluating cellular support for UAVs \cite{3GPP36777}. In this model, all radio links are affected by large-scale fading (comprising antenna gain, path loss, and shadow fading) and small-scale fading.  Among other real-world propagation phenomena, the model accounts for 3D channel directionality, spatially correlated shadow fading, and time-and-frequency correlated small-scale fading. Moreover, all propagation parameters for aerial devices in the model -- such as path loss, probability of LoS, shadow fading, and small-scale fading -- have been derived as a result of numerous measurement campaigns, and explicitly account for the transmitter and receiver heights.

On a given PRB, $\mathbf{h}_{bjk} \in \mathbb{C}^{N_{\mathrm{a}}\times 1}$ denotes the channel vector between BS $b$ and user $k$ in cell $j$. The signal $y_{bk} \in \mathbb{C}$ received by user $k$ in cell $b$ can be expressed as
\begin{equation}
\begin{aligned}
y_{bk} &= \sqrt{P_{\textrm{b}}} \, {\mathbf{h}_{bbk}^{\mathrm{H}} \mathbf{w}_{bk} s_{bk}} + \sqrt{P_{\textrm{b}}} \sum_{i\in\mathcal{K}_b\backslash k} {\mathbf{h}_{bbk}^{\mathrm{H}} \mathbf{w}_{bi} s_{bi}} \\
&+ \! \sqrt{P_{\textrm{b}}} \! \sum_{j \in \mathcal{B} \backslash b} \,\sum_{\,i\in\mathcal{K}_{j}} {\mathbf{h}_{jbk}^{\mathrm{H}} \mathbf{w}_{ji} s_{ji}} + \epsilon_{bk},
\end{aligned}
\label{eqn:rx_signal}
\end{equation}
where $s_{bk} \in \mathbb{C}$ is the unit-variance signal intended for user $k$ in cell $b$, $\epsilon_{bk} \sim \mathcal{CN}(0,\sigma^2_{\epsilon})$ is the thermal noise, and $\mathbf{w}_{bk} \in \mathbb{C}^{N_{\mathrm{a}}\times 1}$ is the transmit precoding employed by BS $b$ to serve user $k$ in cell $b$. The latter is normalized to satisfy the total power constraint, and it is assumed to be a vector of identical scalars for single-user mode operations (representing an analog signal combiner without phase shifters), and a digital zero-forcing (ZF) precoder for multi-user operations, as detailed in Sec.~\ref{sec:SU-MIMO} and Sec.~\ref{sec:MU-MIMO}, respectively. The four terms on the right hand side of (\ref{eqn:rx_signal}) respectively represent: the useful signal, the intra-cell interference from the serving BS (only present for multi-user operations), the inter-cell interference from other BSs, and the thermal noise.

Assuming that the users have perfect channel state information (CSI), the resulting instantaneous signal-to-interference-plus-noise ratio (SINR) $\gamma_{bk}$ at user $k$ in cell $b$ on a given PRB is obtained via an expectation over all symbols, and it is given by
\begin{align}\label{equ:sinr_ue}
\gamma_{bk} = \frac{ P_{\textrm{b}} \, \vert \mathbf{h}_{bbk}^{\mathrm{H}} \mathbf{w}_{bk} \vert^2 }
{P_{\textrm{b}} \!\sum\limits_{i\in\mathcal{K}_b\backslash k} \! \vert \mathbf{h}_{bbk}^{\mathrm{H}} \mathbf{w}_{bi} \vert^2 \!+ 
\! P_{\textrm{b}} \! \sum\limits_{j \in \mathcal{B} \backslash b} \sum\limits_{i\in\mathcal{K}_{j}} \! \vert \mathbf{h}_{jbk}^{\mathrm{H}} \mathbf{w}_{ji} \vert^2 + \sigma^2_{\epsilon}}.
\end{align}
Each SINR value is mapped to the rate delivered on a given PRB by considering ideal link adaptation, i.e., selecting the maximum modulation and coding scheme (MCS) that ensures a desired block error rate (BLER) \cite{3GPP36213}. We set the BLER to $10^{-1}$, which is a sufficiently low value considering that retransmissions further reduce the number of errors. This yields a minimum spectral efficiency of 0.22~b/s/Hz for SINRs between -5.02~dB and -4.12~dB, and a maximum spectral efficiency of 7.44~b/s/Hz for SINRs of 25.87~dB and above. In this paper, we also account for the overhead introduced by control signaling when computing the resultant user rates \cite{3GPP36213}.


\begin{table}
\centering
\caption{System parameters}
\label{table:parameters}
\def\arraystretch{1.2}
\begin{tabulary}{\columnwidth}{ |p{2.4cm} | p{5.55cm} | }
\hline
	\textbf{Deployment} 			&  \\ \hline
  BS distribution				& Three-tier wrapped-around hexagonal grid, 37 sites, three sectors each, one BS per sector \cite{3GPP36777} \\ \hline
  BS inter-site distance 		& 500~m \cite{3GPP36777} \\ \hline
  User distribution 				& 15 users per sector on average \cite{3GPP36777} \\ \hline
	\multirow{2}{*}{GUE distribution} 				& 80\% indoor, horizontal: uniform, vertical: uniform in buildings of four to eight floors  \\ \cline{2-2}
	 				& 20\% outdoor, horizontal: uniform, vertical: 1.5~m \\ \hline
	UAV distribution 				& 100\% outdoor, horizontal: uniform, vertical: uniform between 1.5~m and 300~m \cite{3GPP36777} \\ \hline
	UAVs/GUEs ratio 				& 3GPP Case~3: 7.1\%, Case~4: 25\%, Case~5: 50\% \cite{3GPP36777} \\ \hline
	User association				& Based on RSRP (large-scale fading) \\ \hline \hline
	\textbf{Channel model} 			&  \\ \hline
	Path loss 					& Urban Macro as per \cite{3GPP36777, 3GPP38901}  \\ \hline
	Probability of LoS 					& Urban Macro as per \cite{3GPP36777, 3GPP38901}  \\ \hline
	Shadow fading 		& Urban Macro as per \cite{3GPP36777, 3GPP38901} \\ \hline
	Small-scale fading  	& Urban Macro as per \cite{3GPP36777, 3GPP38901} \\ \hline
	\multirow{2}{*}{Channel estimation}		& Single-user: perfect channel estimation\\ \cline{2-2}
			& Multi-user: UL SRSs with Reuse 3 \\ \hline
	Thermal noise 				& -174 dBm/Hz spectral density \cite{3GPP36777}\\ \hline \hline
	\textbf{PHY} 			&  \\ \hline
	Carrier frequency 		& 2~GHz \cite{3GPP36777} \\ \hline
	System bandwidth 		& 10 MHz with 50 PRBs \cite{3GPP36777} \\ \hline
	BS transmit power 			& 46~dBm \cite{3GPP36777} \\ \hline   
	BS antenna elements 		& Horizontal and vertical half power beamwidth: $65^{\circ}$, max. gain: 8~dBi \cite{3GPP36777} \\ \hline
	BS array configuration 		& Height: $25$~m, mechanical downtilt: $12^{\circ}$, element spacing: $0.5\lambda$ \cite{3GPP36777}\\ \hline
	\multirow{2}{*}{BS array size } 		& Single-user: $8\times 1$ X-POL $\pm 45^{\circ}$, 1 RF chain \\ \cline{2-2}
	  		& Multi-user: $8\times 8$ X-POL $\pm 45^{\circ}$, 128 RF chains  \\ \hline
	\multirow{2}{*}{BS precoder} 		& Single-user: none \\ \cline{2-2}
	  		& Multi-user: zero-forcing  \\ \hline
  \multirow{2}{*}{Power control}		& DL: equal power allocation\\ \cline{2-2}
			& UL: fractional power control with $\alpha = 0.5$, $P_{0} = -58$~dBm, and $P_{\textrm{max}}=23$~dBm \cite{UbeVilRos2008}\\ \hline
	User antenna 		& Omnidirectional with vertical polarization, gain: 0~dBi \cite{3GPP36777} \\ \hline
	Noise figure 			& BS: 7~dB, user: 9~dB \cite{3GPP36777, 3GPP36814} \\ \hline \hline
	\textbf{MAC} 			&  \\ \hline
	Traffic model		& Full buffer \\ \hline
	\multirow{2}{*}{Scheduler}		& Single-user: round-robin \cite{1043857}, one user per PRB \\ \cline{2-2}
			& Multi-user: round-robin, eight users per PRB \\ \hline
\end{tabulary}
\end{table}
\section{Integrating UAVs into Existing Networks}
\label{sec:SU-MIMO}

\begin{figure}[!t]
\centering
\includegraphics[width=0.8\columnwidth]{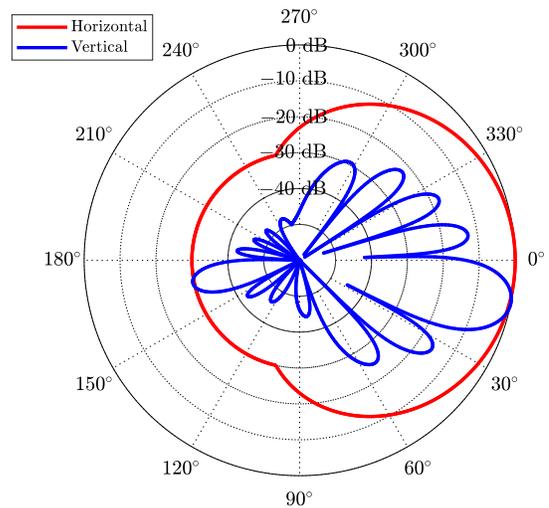}
\caption{Horizontal and vertical antenna pattern (normalized to maximum gain) of a BS consisting of a vertical array of 8 X-POL elements, each with $65^{\circ}$ half power beamwidth, mechanically downtilted by $12^{\circ}$.}
\label{fig:SU-AntennaPattern}
\end{figure}

In this section, we consider a cellular network as depicted in Fig.~\ref{fig:Network}(a), where BSs are equipped with $N_{\mathrm{a}}=16$ antennas arranged in a vertical array of 8 cross-polarized (X-POL) elements, each with $65^{\circ}$ half power beamwidth, mechanically downtilted by $12^{\circ}$ and supported by a single radio-frequency (RF) chain. Such configuration yields the BS antenna pattern depicted in Fig.~\ref{fig:SU-AntennaPattern}. In this setup, each BS serves a maximum of one user on each PRB, without employing digital precoding. We refer to this setup as a \emph{single-user} mode, and we consider it to exemplify existing cellular networks. In single-user mode, equations (\ref{eqn:rx_signal}) and (\ref{equ:sinr_ue}) are simplified as follows: all precoding vectors $\mathbf{w}$ consist of identical scalars, the second term on the right hand side of (\ref{eqn:rx_signal}) vanishes, and so does the first term in the denominator of (\ref{equ:sinr_ue}). For this single-user mode, we will closely examine how UAVs associate to BSs depending on the height of the former, and how the performance of the DL UAV C\&C channel is affected by the UAV height.

\subsection{UAV Association}

Fig.~\ref{fig:SU-DistanceToServer_CDF} shows the cumulative distribution function (CDF) of the 3D distance between a UAV and its best potential serving BS for data transmission and reception. The CDF of said distance is plotted for UAV heights of 1.5~m, 50~m, 75~m, 150~m, and 300~m. For the same set of heights, Fig.~\ref{fig:SU-AntennaGain} depicts the antenna gain between a BS and a UAV aligned to the BS's horizontal bearing as a function of the 2D ground distance between them. By jointly looking at these two figures, three association behaviors can be clearly identified for different UAV heights. 

\subsubsection*{UAVs at 1.5~m}
UAVs at the same height as outdoor GUEs exhibit an association distance smoothly distributed between 35~m and 1~km (Fig.~\ref{fig:SU-DistanceToServer_CDF}). By recalling that the inter-site distance is 500~m, one can make the two following observations:
\begin{itemize}
\item A UAV at 1.5~m falls within the main lobe of a BS as long as its 2D distance exceeds 52~m (Fig.~\ref{fig:SU-AntennaGain}). Moreover, the antenna gain is maximized for 2D distances in the range 80~m-180~m (Fig.~\ref{fig:SU-AntennaGain}), hence associations to BSs other than the closest one are mostly due to shadow fading and LoS conditions.
\item UAVs at 1.5~m generally associate to their closest BS -- i.e., at a distance up to around 250~m (Fig.~\ref{fig:SU-DistanceToServer_CDF}). Association to BSs located at a distance of 500~m or more occurs only in 9\% of the cases (Fig.~\ref{fig:SU-DistanceToServer_CDF}).
\end{itemize}

\subsubsection*{UAVs at 50~m}
For UAVs flying at this moderate height, one can find similarities as well as differences with respect to the previous case of low-height UAVs:
\begin{itemize}
\item For 3D distances of 250~m or less, the CDFs in Fig.~\ref{fig:SU-DistanceToServer_CDF} reveal similar behavior when comparing a UAV at 1.5~m with one at 50~m. Such similarity can be explained from the fact that the vertical distance to a BS -- standing at 25~m -- is similar for the two UAVs. This implies that a BS sees the two UAVs with a similar angle (slightly better for the 1.5~m UAV due to the $12^{\circ}$ BS antenna downtilt), and that the antenna gain as a function of the 2D distance follows a similar pattern (Fig.~\ref{fig:SU-AntennaGain}).
\item For 3D distances of more than 250~m, the two CDFs in Fig.~\ref{fig:SU-DistanceToServer_CDF} exhibit a different behavior. Indeed, UAVs at 50~m are more likely to associate to BSs located 500~m away compared to UAVs at 1.5~m (in 17\% of the cases as opposed to 9\% of the cases). This can be explained by looking at the difference in the two antenna gain trends (Fig.~\ref{fig:SU-AntennaGain}). The higher likelihood of seeing multiple neighboring BSs in LoS for 50~m-high UAVs also contributes to the above phenomenon.
\end{itemize}

\subsubsection*{UAVs at 75~m and above}
\begin{figure}[!t]
\centering
\includegraphics[width=\figwidth]{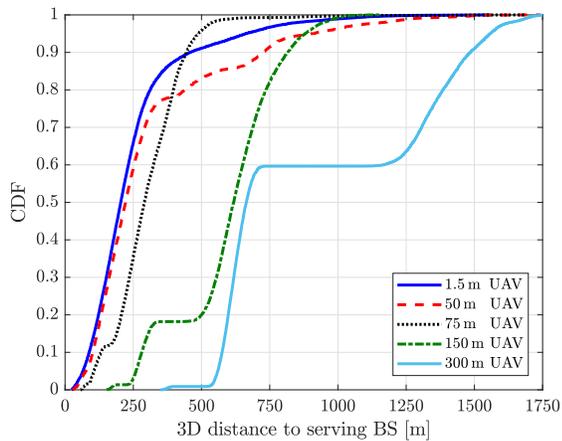}
\caption{CDF of the 3D distance between a UAV and its serving BS in an existing cellular network. Various UAV heights are considered.}
\label{fig:SU-DistanceToServer_CDF}
\end{figure}

\begin{figure}[!t]
\centering
\includegraphics[width=\figwidth]{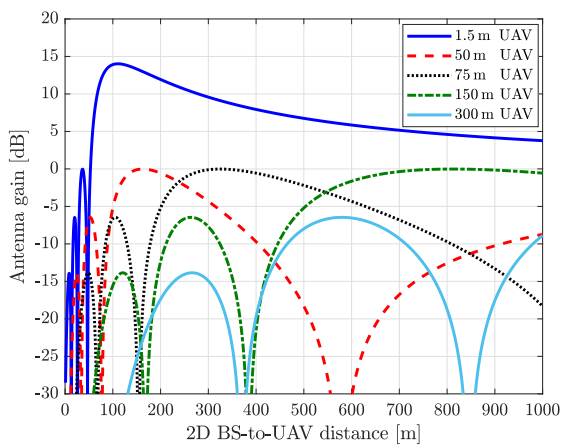}
\caption{Antenna gain between a BS in an existing cellular network and a UAV aligned to the BS's horizontal bearing as a function of their 2D distance. Various UAV heights are considered.}
\label{fig:SU-AntennaGain}
\end{figure}

As a UAV's height increases to 75~m and beyond, not only the association distance grows, but also its CDF exhibits a certain number of steps, each corresponding to a different association distance range (Fig.~\ref{fig:SU-DistanceToServer_CDF}). Such distance ranges are clearly separated from one another, and are due to the secondary lobes of each BS's antenna pattern (see Fig.~\ref{fig:SU-AntennaPattern}). Indeed, when UAVs fly very high, the main BS antenna lobe is only visible at 2D distances larger than 1~km (outside the range of Fig.~\ref{fig:SU-AntennaGain}). The secondary lobes thus play a crucial role in the association process. Owing to an almost free-space propagation, the path loss difference between the closest and the further away BSs is not significant compared to the difference in their respective antenna gains \cite{3GPP36777}. As a result, UAVs tend to associate to BSs located at large distances with better secondary lobe gains.

\begin{figure}[!t]
\centering
\includegraphics[width=\figwidth]{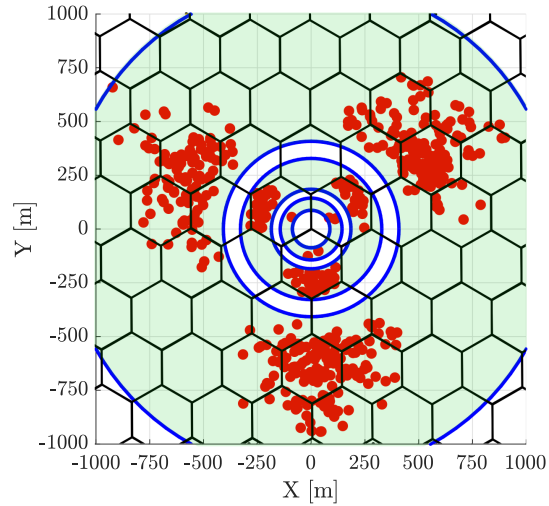}
\caption{2D location samples (red dots) of 150~m-high UAVs associated to a three-sector BS site located at the origin. Distance ranges corresponding to different secondary lobes are shaded and delimited by blue circles.}
\label{fig:SU-Association}
\end{figure}

The above phenomena are further illustrated in Fig.~\ref{fig:SU-Association}. This figure takes the perspective of a three-sector BS located at the origin, and shows samples of the 2D locations of its associated UAVs (red dots) for UAV heights of 150~m. Fig.~\ref{fig:SU-Association} confirms the existence of distance ranges (represented by the green shaded regions), each corresponding to one of the CDF steps in Fig.~\ref{fig:SU-DistanceToServer_CDF} or, equivalently, to one of the secondary lobes in Fig.~\ref{fig:SU-AntennaPattern} and Fig.~\ref{fig:SU-AntennaGain}.

\subsection{UAV C\&C Channel Performance}

\subsubsection*{Carrier signal strength and interference}
\begin{figure}[!t]
\centering
\includegraphics[width=\figwidth]{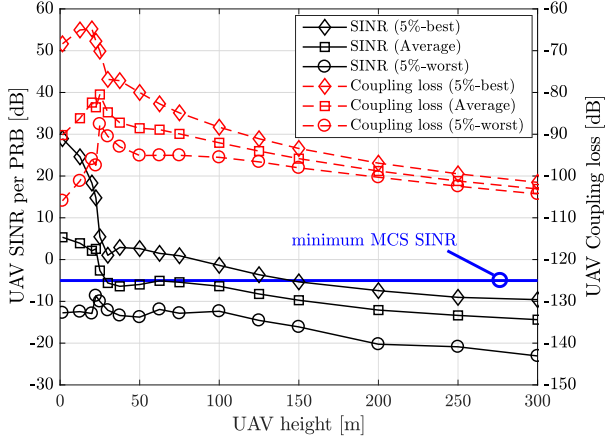}
\caption{Coupling loss (right y-axis) and SINR per PRB (left y-axis) experienced by a UAV as a function of its height in a single-user scenario. The minimum MCS SINR threshold of -5.02~dB is also shown for comparison.}
\label{fig:SU-SINR_CouplingLoss}
\end{figure}

Six curves are plotted in Fig.~\ref{fig:SU-SINR_CouplingLoss} in order to show the coupling loss and the SINR per PRB experienced by a UAV as a function of its height. 
The coupling loss (right y-axis) expresses the carrier signal attenuation between the serving BS and a UAV due to antenna gain, path loss, and shadow fading. On the other hand, the SINR per PRB (left y-axis) also accounts for small-scale fading and for the interference perceived at the UAV. For both metrics, Fig.~\ref{fig:SU-SINR_CouplingLoss} shows the 5\%-best (i.e., 95\%-ile), average, and 5\%-worst values, triggering the following observations:
\begin{itemize}
\item 
As UAVs rise from the ground up to a height of around 25~m, their average coupling loss improves due to closer proximity to the serving BS and increased probability of experiencing a LoS link with the latter. Instead, the 5\%-best UAVs, which were those located in the direction of the main lobe of the BSs, experience a degraded coupling loss as a consequence of a diminished antenna gain. As the UAV height keeps increasing, so does the BS-to-UAV distance, causing the coupling loss to decay.
\item 
While the average UAV coupling loss is moderately improved, a UAV flying at around 25~m generally sees a degraded SINR per PRB. This is caused by the fact that more and more neighboring BSs become visible to the UAV, acting as strong LoS interferers. The opposite occurs for the 5\%-worst UAVs flying at 25~m, which experience a significant improvement in their coupling loss, and as a result, also enhance their SINR. As the UAV height keeps increasing, the SINR keeps decreasing, though more slowly than the coupling loss. This trend is due to a simultaneous slight reduction of the interference as the UAV moves further away from neighboring interfering BSs.
\item Overall, UAVs flying at heights of 25~m and above experience low values of SINR per PRB. In particular, for heights beyond 100~m the average SINR falls below the minimum MCS SINR threshold of -5.02~dB; for heights beyond 150~m even the 5\%-best SINR per PRB falls below said minimum threshold.\footnote{In practice, opportunistic proportional-fair schedulers could be employed that outperform the round-robin scheduler considered in this paper. However, this measure alone would not suffice to bring the UAV C\&C channel performance to an acceptable level in single-user mode networks. Indeed, Fig.~\ref{fig:SU-SINR_CouplingLoss} shows that even the 5\%-best UAVs, which corresponds to those with large channel fading gains, experience very low values of SINR.} 
\end{itemize}

\subsubsection*{C\&C channel data rates}
\begin{figure}[!t]
\centering
\includegraphics[width=\figwidth]{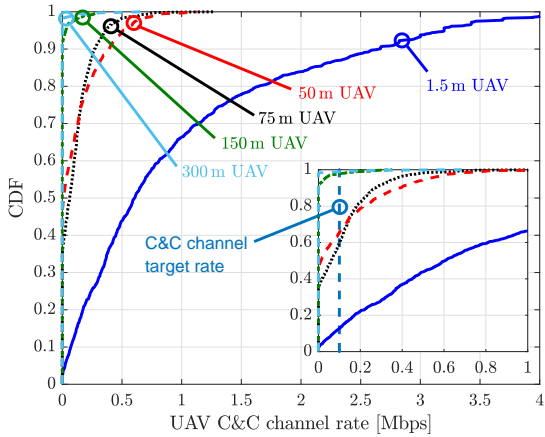}
\caption{CDF of the UAV C\&C channel rates as a function of the UAV height in a single-user scenario. In the enlargement, the target rate of 100~kbps is also shown for comparison.}
\label{fig:SU-McsRate_CDF}
\end{figure}

The measured values of SINR per PRB can be translated into the data rate performance of the UAV C\&C channel over a 10~MHz bandwidth, as per the MCS selection described in Sec.~\ref{sec:System_Model}. Fig.~\ref{fig:SU-McsRate_CDF} shows said performance for various UAV heights, motivating the following conclusions:
\begin{itemize}
\item 
UAVs flying at a height of 1.5~m achieve the target rate of 100~kbps 87\% of the time. Moreover, 34\% of the time their data rates even exceed 1~Mbps.
\item 
UAVs flying at around 50~m and 75~m only achieve the target rate 35\% and 40\% of the time, respectively. The achievable rates for this range of UAV heights almost never reach 1~Mbps (0.3\% of the time).
\item As UAVs fly higher, the target rate of 100~kbps can only be achieved for a small fraction of time, amounting to just 2\% and 1\% for heights of 150~m and 300~m, respectively.
\end{itemize}
The above results allow us to conclude that cellular networks with heavy data traffic and that simply rely on BS sectorization and single-user mode operations are unlikely to be able to support the much-needed C\&C channel for UAVs flying at reasonable heights.\footnote{While there is a strong interest in reusing the existing network infrastructure with downtilted antenna arrays, dedicated BSs with uptilted arrays could be considered in the future to provide C\&C and connectivity just to UAVs.}

\section{Supporting UAVs through Massive MIMO}
\label{sec:MU-MIMO}

In this section, we consider a cellular network as depicted in Fig.~\ref{fig:Network}(b), where BSs are equipped with massive MIMO antenna arrays and provided with beamforming and spatial multiplexing capabilities. In particular, we consider $N_{\mathrm{a}}=128$ antennas, arranged in an $8\times 8$ planar array of $\pm 45^{\circ}$ cross-polarized elements, supported by 128 RF chains. Since massive MIMO allows the transmission of beamformed control channels, users tend to associate to a nearby BS. We allow each BS $b$ to serve a maximum of $K_b=8$ users per PRB through digital ZF precoding. We denote this setup as a \emph{multi-user} mode scenario, and we consider it to exemplify next-generation massive MIMO cellular deployments. In such multi-user mode setting, the network operates in a time-division duplexing (TDD) fashion, where channels are estimated at the BS through the use of UL sounding reference signals (SRSs) -- commonly known as \emph{pilots} -- sent by the users under the assumption of channel reciprocity \cite{MarLarYanBook2016}. In the following, we describe in detail the sequence of operations performed in this multi-user system.

\subsubsection*{Channel state information acquisition}
Let the pilot signals span $M_{\mathrm{p}}$ symbols. The pilot transmitted by user $k$ in cell $b$ is denoted by $\mathbf{v}_{\textrm{i}_{bk}} \in \mathbb{C}^{M_{\mathrm{p}}}$, where $\textrm{i}_{bk}$ is the index in the pilot codebook, and all pilots in the codebook form an orthonormal basis \cite{MarLarYanBook2016}. Each pilot signal received at the BS suffers from \emph{contamination} due to pilot reuse across cells. We assume pilot Reuse 3, i.e., the set of pilot signals is orthogonal among the three $120^{\circ}$ BS sectors of the same site, but it is reused among all BS sites, generating contamination. This solution is particularly practical from an implementation standpoint, since it involves coordination only between the three co-located BSs of the same BS deployment site. Each BS sector randomly allocates its pool of pilots to its served users. The collective received signal at BS $b$ is denoted as $\mathbf{Y}_b \in \mathbb{C}^{N_{\mathrm{a}} \times M_{\mathrm{p}}}$, and given by
\begin{equation}
\mathbf{Y}_b = \sum_{j \in \mathcal{B}} \sum_{k\in\mathcal{K}_j} \sqrt{P_{jk}} \mathbf{h}_{bjk} \mathbf{v}_{\mathrm{i}_{jk}}^{\textrm{T}} + \mathbf{N}_b,	
\label{eqn:Yb}
\end{equation}
where $\mathbf{N}_b$ contains the additive noise at BS $b$ during pilot signaling with independent and identically distributed entries following $\mathcal{CN}(0,\sigma^2_{\epsilon})$, and $P_{jk}$ is the power transmitted by user $k$ in cell $j$. We assume fractional UL power control as follows \cite{UbeVilRos2008}
\begin{equation}
P_{jk} = \min\left\{ P_{\textrm{max}}, P_0 \cdot \bar{h}_{jjk}^\alpha \right\},
\label{eqn:power_control}
\end{equation}
where $P_{\textrm{max}}$ is the maximum user transmit power, $P_0$ is a cell-specific parameter, $\alpha$ is a path loss compensation factor, and $\bar{h}_{jjk}$ is the average channel gain measured at UE $k$ in cell $j$ based on the RSRP~\cite{3GPP36201,3GPP36213}. The aim of (\ref{eqn:power_control}) is to compensate only for a fraction $\alpha$ of the path loss, up to a limit imposed by $P_{\textrm{max}}$.

The received signal $\mathbf{Y}_b$ in (\ref{eqn:Yb}) is processed at BS $b$ by correlating it with the known pilot signal $\mathbf{v}_{\textrm{i}_{bk}}$, thus rejecting interference from other orthogonal pilots. BS $b$ hence obtains the following least-squares channel estimate for user $k$ in cell $b$~\cite{kay}
\begin{equation}
\begin{aligned}
\widehat{\mathbf{h}}_{bbk} &=
\frac{1}{\sqrt{P_{bk}}} \mathbf{Y}_b \mathbf{v}_{\mathrm{i}_{bk}}^{*} = \mathbf{h}_{bbk} \\
& \enspace+ \frac{1}{\sqrt{P_{bk}}} \Big( \sum_{j \in \mathcal{B} \backslash b} \,\sum_{k\in\mathcal{K}_j} \sqrt{P_{jk}} \mathbf{h}_{ijk} \mathbf{v}_{\mathrm{i}_{jk}}^{\textrm{T}} +
\mathbf{N}_i \Big) \mathbf{v}_{\mathrm{i}_{bk}}^{*}
\label{eqn:PC}
\end{aligned}
\end{equation}
where intra-cell pilot contamination is not present since BS $b$ allocates different pilots for the users in its own cell.

\subsubsection*{Spatial multiplexing through digital precoding}
Each BS simultaneously serves multiple users on each PRB through ZF precoding, attempting to suppress all intra-cell interference. Let us define the estimated channel matrix $\widehat{\mathbf{H}}_b \in \mathbb{C}^{N_{\mathrm{a}}\times K_b}$ as
\begin{equation}
\widehat{\mathbf{H}}_b = \left[ {\widehat{\mathbf{h}}_{bb1}},\ldots,{\widehat{\mathbf{h}}_{bb{K_b}}} \right].
\label{eqn:Hb}
\end{equation}
The ZF precoder
\begin{equation}
	\mathbf{W}_b = \left[ {\mathbf{w}}_{b1},\ldots,{\mathbf{w}}_{b K_{b}} \right]	
\label{eqn:precoder}
\end{equation}
at BS $b$ can be calculated as \cite{SpeSwiHaa:04}
\begin{equation}
	\mathbf{W}_b = \widehat{\mathbf{H}}_b \left( \widehat{\mathbf{H}}_b^{\mathrm{H}} \widehat{\mathbf{H}}_b \right)^{-1} \left(\mathbf{D}_b\right)^{-\frac{1}{2}},
	\label{eqn:ZF}
\end{equation}
where the diagonal matrix $\mathbf{D}_b$ is chosen to meet the transmit power constraint with equal user power allocation\footnote{More sophisticated power allocation policies -- e.g., max-min SINR -- may be considered to further improve the performance of the worst UAVs.}, i.e., $\Vert \mathbf{w}_{bk}\Vert^2 = P_{\mathrm{b}}/K_{b}$ $\forall k, b$. The SINR on a given PRB for user $k$ can be calculated from (\ref{equ:sinr_ue}), with the precoding vectors $\mathbf{w}_{bk}$ obtained as in (\ref{eqn:precoder}).

In the remainder of this section, we will \emph{A)} evaluate the UAV C\&C channel performance improvement achieved through massive MIMO, and \emph{B)} study what the presence of UAVs entails for the GUEs performance.

\subsection{UAV C\&C Channel Performance Improvement}

Similarly to Fig.~\ref{fig:SU-SINR_CouplingLoss} and Fig.~\ref{fig:SU-McsRate_CDF} for the single-user mode case, we now show the coupling loss (Fig.~\ref{fig:MU-SINR_CouplingLoss}, right y-axis), SINR per PRB (Fig.~\ref{fig:MU-SINR_CouplingLoss}, left y-axis), and C\&C data rate (Fig.~\ref{fig:MU-McsRate_CDF}) experienced by a UAV as a function of its height in a multi-user massive MIMO setup. Both Fig.~\ref{fig:MU-SINR_CouplingLoss} and Fig.~\ref{fig:MU-McsRate_CDF} consider the 3GPP Case 3, i.e., one UAV and 14 GUEs per sector \cite{3GPP36777}. In order to evaluate the gains theoretically achievable with multi-user massive MIMO, in these figures perfect CSI is assumed available at the BSs, i.e., no pilot contamination is accounted for. A more realistic channel estimation through SRSs as in (\ref{eqn:PC}) and the effect of pilot contamination on the performance of both UAVs and GUEs will be thoroughly discussed in Sec.~\ref{subsec:MU-MIMO_CSI}.

\subsubsection*{Carrier signal strength and interference}
\begin{figure}[!t]
\centering
\includegraphics[width=\figwidth]{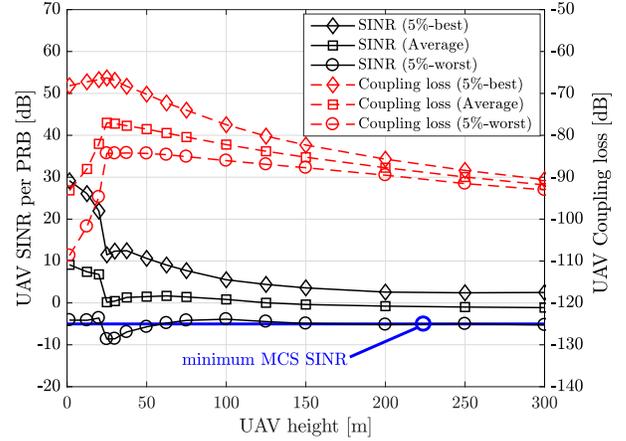}
\caption{Coupling loss (right y-axis) and SINR per PRB (left y-axis) experienced by a UAV as a function of its height in a multi-user scenario with perfect CSI (Case 3). The minimum MCS SINR threshold of -5.02~dB is also shown as a benchmark.}
\label{fig:MU-SINR_CouplingLoss}
\end{figure}

Comparing Fig.~\ref{fig:MU-SINR_CouplingLoss} to Fig.~\ref{fig:SU-SINR_CouplingLoss} provides the following insights:
\begin{itemize}
\item
Consistently with Fig.~\ref{fig:SU-SINR_CouplingLoss}, a UAV flying at around 25~m generally sees an improved coupling loss but a degraded SINR per PRB. This is due to the fact that more neighboring BSs become visible to the UAV, acting as strong LoS interferers.
\item 
Employing massive MIMO at the BSs while keeping the same mechanical downtilt improves the UAVs' coupling loss, which is measured at the output of the first RF chain \cite{3GPP38901}, thanks to an increased antenna gain towards the sky.
\item 
The SINR per PRB experienced by a UAV is largely improved in a massive MIMO system, owing to two phenomena. First, UAVs benefits from a beamforming gain from the serving BS, which can now send beams into the sky as well. Second, since most users are GUEs, neighboring BSs tend to point most of their beams downwards, greatly reducing the interference generated at the UAVs.
\item 
Overall, most UAVs experience values of SINR per PRB above the minimum MCS threshold. In particular, the average SINR per PRB is well above said threshold for any UAV height. Moreover, even the 5\%-worst UAVs meet the minimum SINR threshold for most UAV heights.
\end{itemize}

\subsubsection*{C\&C channel data rates}
\begin{figure}[!t]
\centering
\includegraphics[width=\figwidth]{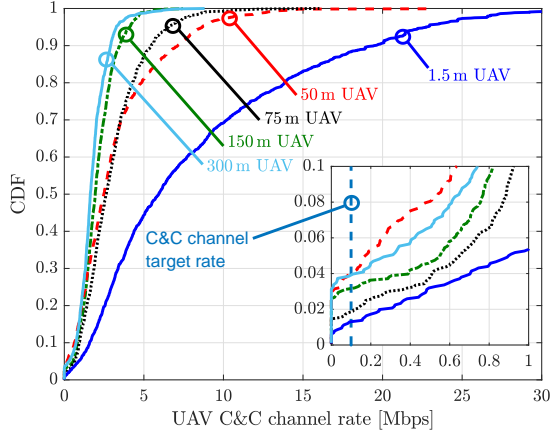}
\caption{CDF of the UAV C\&C channel rates as a function of the UAV height in a multi-user scenario with perfect CSI (Case 3). In the enlargement, the target rate of 100~kbps is also shown as a benchmark.}
\label{fig:MU-McsRate_CDF}
\end{figure}

Fig.~\ref{fig:MU-McsRate_CDF} shows the data rate performance of the UAV C\&C channel in a multi-user massive MIMO setup for various UAV heights. Comparing this figure to Fig.~\ref{fig:SU-McsRate_CDF} provides the reader with a key takeaway: unlike single-user mode cellular networks, massive MIMO networks have the potential to support a 100 kbps UAV C\&C channel with good reliability, namely in at least 96\% of the cases for all UAV heights under consideration. Indeed, the data rates in a massive MIMO network are largely improved owing to both an SINR gain (as per Fig.~\ref{fig:MU-SINR_CouplingLoss}) and a spatial multiplexing gain provided by the fact that eight users, between UAVs and GUEs, are simultaneously allocated the same PRB.

\subsection{Performance Interplay between UAVs and Ground Users}\label{subsec:MU-MIMO_CSI}

We now study how supporting the UAV C\&C channel through cellular networks may affect the performance of GUEs. In particular, we evaluate the impact of UAVs in both single-user and multi-user mode settings with the user height distributions specified in Table~\ref{table:parameters}. For the latter, we discuss the impact of SRS reuse and contamination, as well as the impact of UL power control in the channel estimation phase. 

\subsubsection*{3GPP case studies}
\begin{figure}[!t]
\centering
\includegraphics[width=\figwidth]{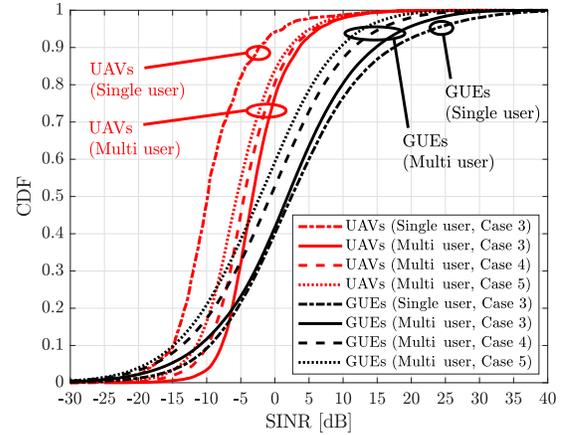}
\caption{SINR per PRB experienced by UAVs and GUEs in single-user and multi-user scenarios with SRS Reuse 3 and UL fractional power control (various 3GPP cases).}
\label{fig:SU-MU-SINR}
\end{figure}

Fig.~\ref{fig:SU-MU-SINR} shows the SINR per PRB for both UAVs and GUEs in the presence of realistic CSI acquisition with SRS Reuse 3 and UL fractional power control. The figure considers the 3GPP Cases 3, 4, and 5, corresponding to one UAV and 14 GUEs, three UAVs and 12 GUEs, and five UAVs and 10 GUEs per sector, respectively. Fig.~\ref{fig:SU-MU-SINR} carries multiple consequential messages:
\begin{itemize}
\item 
In spite of an imperfect CSI available at the BSs, the UAV SINR per PRB greatly improves when moving from a single-user to a multi-user mode scenario. This is due to a beamforming gain paired with a reduced interference from neighboring BSs that focus most of their energy downwards.
\item 
In line with the above, the UAV SINR per PRB in multi-user mode scenarios is reduced when moving from Case 3 to Cases 4 and 5, mainly because \emph{(i)} a larger number of UAVs leads to an increased CSI pilot contamination due to their strong LoS channel to many BSs, and \emph{(ii)} neighboring cells point more beams upwards, thus generating more inter-cell interference at the UAVs. On the other hand -- although not explicitly shown in the figure -- the number of UAVs does not affect the SINR in single-user mode scenarios.
\item 
Unlike the UAV SINR, the GUE SINR does not improve when moving from a single-user to a multi-user mode scenario. This is mainly due to the severe pilot contamination incurred by GUEs, which outweighs any beamforming gains. Indeed, each GUE's SRS is likely to collide with the SRS of at least one UAV in a neighboring cell in the scenario considered, with said UAV being likely to experience a strong LoS link with the GUE's serving BS.
\item 
Accordingly, the GUE SINR per PRB further degrades when moving from Case 3 to Cases 4 and 5, since the presence of more UAVs in neighboring cells causes the pilot contamination effect to increase its severity.  
\end{itemize}

\subsubsection*{Benefits of massive MIMO and UL power control}
\begin{figure}[!t]
\centering
\includegraphics[width=\figwidth]{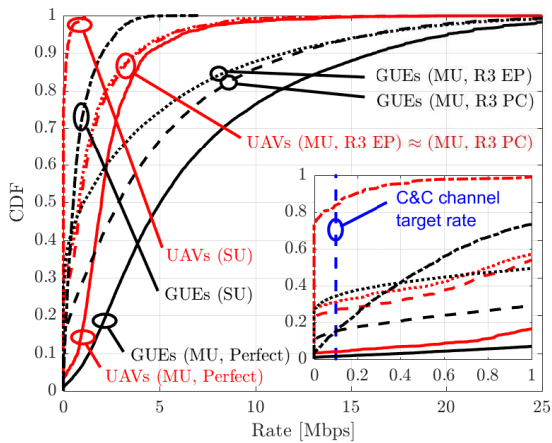}
\caption{Rates achieved by UAVs and GUEs in multi-user scenarios under: \emph{(i)} perfect CSI -- ``MU, Perfect'' (solid), \emph{(ii)} SRS Reuse 3 and UL fractional power control -- ``MU, R3 PC'' (dashed), and \emph{(iii)} SRS Reuse 3 with equal power -- ``MU, R3 EP'' (dotted).
The figure also shows the rates in a single-user scenario -- ``SU'' (dash-dot) and, in the enlargement, the UAV C\&C target rate of 100~kbps.}
\label{fig:MU-McsRate_CDF_CSI}
\end{figure}

The ultimate DL rate performance achievable by UAVs and GUEs is shown in Fig.~\ref{fig:MU-McsRate_CDF_CSI} for the 3GPP Case 3, i.e., one UAV and 14 GUEs per sector. This figure not only illustrates the gains provided by multi-user massive MIMO networks, but it also highlights the crucial role played by UL power control for CSI acquisition through a comparison of three scenarios: \emph{(i)} perfect CSI (``Perfect''), \emph{(ii)} imperfect CSI obtained through SRS Reuse 3 and fractional UL power control (``R3 PC''), and \emph{(iii)} imperfect CSI obtained through SRS Reuse 3 and equal UL power allocation (``R3 EP''). Fig.~\ref{fig:MU-McsRate_CDF_CSI} allows us to conclude this section with the following key takeaways:
\begin{itemize}
\item 
Pilot contamination can significantly degrade the rate performance of both UAVs and GUEs. Indeed, the median UAV rates attained with imperfect CSI acquisition are reduced to 40\% of those achievable without channel estimation errors.
\item 
UL fractional power control does not significantly help to protect the UAV C\&C channel. This is because UAVs generally have a strong LoS channel to a large number of BSs, which entails that they are the main source of pilot contamination. As a result, UAVs do not undergo substantial performance gains with UL power control because both their signal and interference powers are similarly reduced. Instead, UL power control is a tremendously helpful technique for GUEs severely affected by pilot contamination, which are those located in the lower tail of the CDF curve. These GUEs benefit from the large power reduction of the UAV-generated SRSs against their more conservative power adjustment.
\item 
Massive MIMO boosts the GUEs' data rates. This is owed to the multiplexing gain rather than to SINR gain, as illustrated by the non-improving SINRs in Fig.~\ref{fig:SU-MU-SINR}. As for the UAV C\&C channel, massive MIMO is a key enabler, achieving the target rate of 100~kbps in 74\% of the cases even under pilot contamination (``MU, R3 PC'').
\item 
Availing of massive MIMO with perfect CSI would allow to meet said C\&C channel target rate in 96\% of the cases, as opposed to a mere 16\% under single-user setups. In order to close the performance gap caused by pilot contamination, one may resort to better pilot assignment and more advanced channel estimation and precoding techniques based on multi-cell processing \cite{BjoHoySanBook2017,BjoHoySan2018, 6415397}. These schemes leverage in fact channel directionality, which invariably occurs in BS-to-UAV links.
\end{itemize}

\section{Future Outlook}
\label{sec:future}

\begin{table*}[t]
\centering
\caption{Summary of complementary solutions to improve UAV C\&C channel support and UAV-GUE interplay}
\label{table:solutions}
\renewcommand{\arraystretch}{1.5}
\begin{tabular}{|c|c|p{4.85cm}|p{5.4cm}|}\hline
\multicolumn{1}{|c|}{\textbf{Solution}} & \multicolumn{1}{c|}{\textbf{Domain}} & \multicolumn{1}{c|}{\textbf{Approach}} & \multicolumn{1}{c|}{\textbf{Challenges/drawbacks}} \\\hline
Interference blanking & Time and frequency & Neighboring cells relieve UAVs of DL interference by employing ABSs & Less efficient for high-height UAVs, since they require blanking from large cell clusters \\\hline
Opportunistic scheduling & Time and frequency & Neighboring cells schedule their respective UAVs of different PRBs & Scheduling coordination among cells; less viable for high density of UAVs \\\hline
Fractional pilot reuse & Time, frequency, and space & Conservative pilot reuse for UAVs, more aggressive pilot reuse for GUEs & SRS coordination among cells; larger pilot overhead for high UAV densities \\\hline
Uplink power control & Power & Customize $P_{0}$ and $\alpha$ in (\ref{eqn:power_control}), assigning lower values to UAVs than to GUEs & RSRP from many cells may vary fast with high UAV mobility \\\hline
Cooperative MIMO & Space & Suppress interference (CEA precoding) or turn it into useful signal (CoMP) & SRS coordination (CEA precoding); X2 interface between many cells (CoMP) \\\hline
UAV directional antennas & Space & UAV DL C\&C signal is strengthened,  UL interference to GUEs is reduced & Detrimental for UL of UAVs unless beamforming or antenna selection are possible \\ \hline
\end{tabular}
\end{table*}

While massive MIMO provides substantial improvements to UAV cellular communications, one may also use complementary techniques to further improve the performance of UAVs as well as their interplay with traditional GUEs. Table~\ref{table:solutions} summarizes what we consider to be the most promising solutions, worthy of future research. The remainder of this section is dedicated to an overview of their potentials and challenges. We note that most of the solutions listed rely on the ability of identifying a UAV, which can be accomplished either: \emph{(i)} through mobility and handover history; or \emph{(ii)} with the help of the UAV itself via enhanced measurement report, in-flight mode indication, or altitude information messages \cite{3GPP36777}.

\subsubsection*{Interference blanking}

Having understood that the UAV C\&C channel performance bottleneck is due to inter-cell interference from a large number of neighboring cells, rather than to a weak carrier signal, one may consider silencing the strongest interfering cells. To this end, the set of strongest interferers could use almost blank subframes (ABSs) on the time-frequency resources that have been assigned to a UAV C\&C channel, thus guaranteeing that the latter experiences a satisfactorily high SINR \cite{R11717287,R11718267}. Determining the sets of BSs that are to use ABSs can have a significant impact on the resultant performance of both UAVs and GUEs. As discussed in Sec.~\ref{sec:SU-MIMO}, the height of a UAV determines the number of strong interfering cells. As a result, the higher the UAV, the larger the number of cells that should protect the UAV C\&C channel by undergoing silent phases. This may pose a problem in terms of both the size of the BS cluster that needs to be coordinated and the amount of time-frequency resources that must be sacrificed to protect each UAV C\&C channel. As a result, this solution may only be suitable for cellular networks with a low density of UAVs. 

\subsubsection*{Opportunistic scheduling}

In a massive MIMO network, a more efficient alternative to blanking could be scheduling UAVs on different PRBs. Indeed, part of the benefits brought by massive MIMO to the UAV C\&C channel are due to the fact that the neighboring cells of each UAV point most their beams to GUEs, thus focusing most of their radiated power downwards. This can be observed in Fig.~\ref{fig:SU-MU-SINR}, where reducing the density of UAVs by a factor of five -- i.e., moving from Case~5 to Case~3 -- increases by 50\% the number of UAVs that achieve the minimum MCS SINR. Such phenomenon suggests that close-by BSs could opportunistically schedule the DL C\&C channel of their UAVs on different PRBs, making sure that UAVs are not interfered by other beams pointing upwards in their vicinity.
 

\subsubsection*{Fractional pilot reuse}

Both UAVs and GUEs served through a massive MIMO network see their rates increased by orders of magnitude when compared to systems where BSs have a limited number of antennas. Said conclusion holds for networks that employ pilot Reuse~3, and we also showed that these gains can be further boosted if BSs avail of perfect CSI. While allocation of fully orthogonal pilots across the network provides a higher quality CSI, the associated overhead makes this approach infeasible in practical systems. As a trade-off between conservative and aggressive reuse approaches, fractional reuse could be the most suitable strategy in a network that accommodates UAVs with strong LoS links to a plurality of BSs \cite{GalCamLopWCNC2017}. Indeed, each BS could selflessly relieve neighboring BSs of severe pilot contamination by assigning a dedicated pool of pilots to its served UAVs, where such pool is agreed beforehand and reused sporadically, e.g., with a reuse factor larger than three. At the same time, each BS could adopt a more aggressive reuse for all remaining pilots, which are assigned to GUEs. As a result, both GUEs and UAVs would be relieved of most pilot contamination, without incurring a large overhead.

\subsubsection*{Uplink power control}

While we showed in Fig.~\ref{fig:MU-McsRate_CDF_CSI} that the effect of pilot contamination can also be alleviated through conventional fractional UL power control, one may also think of more effective techniques. A possible improvement could be achieved by generalizing (\ref{eqn:power_control}) with customized values of $P_{0}$ and $\alpha$, e.g., assigning lower values to UAVs than to GUEs, or even accounting for the specific UAV height \cite{YajWanGao2018}. A further generalization would involve modifying the approach to account not only for the RSRP from the serving cell but also for the RSRP from neighboring cells. While (\ref{eqn:power_control}) is thought for a GUE -- increasing the transmit power as the GUE's RSRP from the serving BS decreases -- this approach may not be always suitable for a UAV. Indeed, when considering a UAV, a decreasing RSRP may be a symptom of a high height. In this case, increasing the power as per (\ref{eqn:power_control}) would exacerbate the interference generated to a plurality of BSs in LoS. Adapting the power control formula in (\ref{eqn:power_control}) to account for the RSRP from multiple cells could solve this problem, e.g., by forcing a UAV to reduce its power when the RSRPs from both serving and neighboring cells are low and similar.

\subsubsection*{Cooperative MIMO}

One could resort to multi-cell signal processing to boost the UAV C\&C channel SINR. In cooperative multipoint (CoMP), a cluster of BSs coherently transmit towards each UAV, aiming to turn interference into useful signal. Despite the promises in terms of rate improvements, CoMP poses a significant overhead over the BS-to-BS X2 interface due to the need of sharing UAV data and achieving a tight symbol-level synchronization \cite{R117011313,LozHeaAnd:13}. As a more practical alternative, cell-edge-aware (CEA) precoding techniques exploiting inter-cell CSI -- acquired through coordinated orthogonal SRSs -- may be adopted to steer interference towards the channel nullspace of neighboring UAVs \cite{YanGerQueTSP2016,GerGarLop2016,GarGerGal2017}. While known to be effective for interference management at cell-edge GUEs \cite{GerGarLopWCNC2017}, both CoMP and CEA precoding may face a number of challenges in UAV setups, where the number of sites that must be coordinated grows due to the larger number of interfering cells \cite{LinYajMur2018}, and where the high UAV mobility may entail frequent updates of the coordination clusters.

\subsubsection*{UAV directional antennas}

While most research efforts have focused on solutions to be implemented at the BS, the 3GPP community is also exploring the possibility of tackling the interference problem directly at the user side, by equipping UAVs with directional antennas. Preliminary contributions have shown that simply implementing a directional antenna without beamsteering capabilities does not provide significant benefits to the DL UAV performance \cite{R11714071}. Moreover, such setup may harm the UL UAV performance, due to the high probability of not pointing the antenna towards the most adequate BS. More encouraging studies argue that antenna selection or beamforming capabilities at the UAV can both enhance the DL C\&C signal reception and mitigate the UL interference it generates towards other BSs \cite{NguAmoWig2018}. However, the latter solution requires increasing the UAV hardware and computational complexity, and it is unclear whether all manufacturers will be willing to implement it.
\section{Conclusions}
\label{sec:conclusions}

In this article, we took a fresh look at UAV cellular communications, following the most recent trends from the industry, academia, and the standardization fora. Employing realistic 3GPP channel and system models, we evaluated the performance of the downlink command and control channel when supported by either: a traditional cellular network serving one user per transmission time interval, or a multi-user massive MIMO network exploiting spatial multiplexing. Besides comparing the capability and reliability of existing cellular infrastructure to next-generation deployments, we closely examined how aerial users of different heights undergo dissimilar cell selection, carrier signal interference, and pilot contamination. We concluded by discussing complementary procedures that leverage the time, frequency, power, and spatial domains to further enhance UAV cellular communications, and that we believe merit further investigation.
\balance
\ifCLASSOPTIONcaptionsoff
  \newpage
\fi
\bibliographystyle{IEEEtran}
\bibliography{Strings_Gio,Bib_Gio}
\end{document}